\documentclass[]{aa}

\usepackage{graphicx}             
\usepackage{float}                 
\usepackage{placeins}              
\usepackage{adjustbox}             
\usepackage{lscape}                
\usepackage{wrapfig}               
\usepackage{epstopdf}              
\usepackage{subfigure}             
\usepackage{booktabs}              
\usepackage{amsmath}               
\usepackage[T1]{fontenc}           
\usepackage{float}                 
\usepackage{gensymb}               
\usepackage{rotating}              
\usepackage{siunitx}
\usepackage{dcolumn}
\usepackage{hyperref}

\usepackage{txfonts}
\usepackage{natbib}
\usepackage{graphicx}  
\usepackage{amsmath}   
\usepackage{multirow}  
\usepackage{pdflscape}
\bibpunct{(}{)}{;}{a}{}{,} 

\DeclareRobustCommand{\VAN}[3]{#2}
\let\VANthebibliography\thebibliography
\def\thebibliography{\DeclareRobustCommand{\VAN}[3]{##3}\VANthebibliography}

\DeclareGraphicsExtensions{.ps}
\usepackage{graphicx}
\usepackage{txfonts}

\begin{document}

   \title{Spectroscopic modeling of ionic structure in stellar winds of high-mass X-ray binaries}
   \subtitle{}

\author{
G. Sanjurjo-Ferrín\inst{1}
\and V. Grinberg\inst{2}
\and I. El Mellah\inst{3}
\and J.M. Torrejón\inst{1}
\and R. Ballhausen\inst{4,5}
\and J. Planelles-Villalba\inst{1}
\and S. Reyero-Serrantes\inst{2}
\and J.J. Rodes-Roca\inst{1}
\and M. Martínez-Chicharro\inst{6}
}

\institute{
Instituto Universitario de Física Aplicada a las Ciencias y las Tecnologías, Universidad de Alicante, Ap. Correos 99, E-03080 Alicante, Spain
\and
European Space Agency (ESA), European Space Research and Technology Centre (ESTEC), Keplerlaan 1, 2201 AZ Noordwijk, The Netherlands.
\and
Departament de Física, EEBE, Universitat Politècnica de Catalunya, c/Eduard Maristany 16, 08019 Barcelona, Spain
\and
Department of Astronomy, University of Maryland, College Park, MD 20742, USA
\and
NASA Goddard Space Flight Center, Astrophysics Science Division, Greenbelt, MD 20771, USA
\and
Departamento Didáctica Ciencias Experimentales, Sociales y Matemáticas Universidad Complutense.
}

\abstract
{High-mass X-ray binaries (HMXBs) provide a natural laboratory to study radiatively driven stellar winds under strong external X-ray irradiation. As the compact object moves along its orbit, the wind density and ionization structure vary with orbital phase, leaving characteristic signatures in X-ray emission and absorption features. The amplitude and morphology of this variability depend on the system geometry, including the orbital inclination (via line-of-sight and occultation effects) and the orbital eccentricity (via phase-dependent changes in the orbital separation).}{We present a computational framework that connects phase-resolved spectroscopic variability to the three-dimensional structure of irradiated winds, and that enables inference of physically meaningful wind--irradiation parameters within a Bayesian setting.}{We combine photoionization calculations with a parametric wind description and orbital geometry to construct three-dimensional maps of density and ionization. From these maps we compute orbital-phase-dependent diagnostics, accounting for geometric occultation and wind inhomogeneity through a clumping prescription. We then use Bayesian inference to compare model predictions with phase-resolved observables and to quantify parameter constraints and degeneracies.}{The framework reproduces the main orbital-phase-dependent trends expected for irradiated winds and yields robust constraints on the system ionization balance. While combinations of wind and luminosity parameters are well constrained, individual parameters can remain partially degenerate depending on the orbital configuration and data quality.}{This modular and computationally efficient approach provides a route to interpret HMXB phase variability in physical terms, and offers a foundation for future extensions toward forward spectral modeling and population-level applications.}


   \maketitle

\section{Introduction}

High-mass X-ray binaries (HMXBs), including the subclass of supergiant systems (SgXBs), are composed of a compact object---a neutron star (NS) or a black hole (BH)---and an O- or B-type donor with mass $\gtrsim 10\,\mathrm{M_\odot}$. In many HMXBs, the compact object follows a close orbit and is embedded in the donor’s stellar wind. X-ray emission is powered by accretion, through the conversion of gravitational potential energy released by matter falling onto the compact object (see, e.g., \citealt{frank_king_raine_2002}). In SgXBs, wind accretion is typically the dominant mass-transfer channel, and the wind properties directly shape the emergent X-ray spectrum. After being produced near the compact object, the X-ray flux propagates through the wind and interacts with the surrounding material, giving rise to spectral features such as emission and absorption lines \citep[for a review see][]{Martinez-Nunez_2017}.These systems also serve as crucial environments for studying evolutionary processes, as they can act as progenitors of compact-object mergers that are subsequently observed via gravitational-wave signals (see \cite{2012arXiv1208.2422B}).

In massive-star HMXBs, the donor winds are line-driven in the sense of the Castor, Abbott \& Klein (CAK; \citealt{1975ApJ...195..157C}) mechanism, where radiation pressure on ultraviolet metal lines accelerates the outflow. Line-driven winds are intrinsically unstable, leading to shocks and small-scale inhomogeneities commonly referred to as clumps \citep{1984ApJ...284..337O,1988ApJ...335..914O}. Using hydrodynamical simulations, \citet{2012MNRAS.421.2820O} showed that wind structure induces pronounced variability, including strong changes in the predicted accretion-powered X-ray light curves. Their models describe dense shells separated by rarefied regions, with density fluctuations spanning many orders of magnitude. In the broader massive-star context, \citet{2008A&ARv..16..209P} discuss wind clumping primarily in terms of clumping factors and volume filling factors, rather than a unique finite clump to inter-clump density ratio. In that framework, the inter-clump medium is often assumed to be void, so the implied clump to inter-clump density contrast is formally very large. 

While the line-driven instability predicts stronger fragmentation at larger radii, observational evidence indicates that clumping may also be present close to the photosphere \citep[e.g.,][]{2013MNRAS.428.1837S,2015ApJ...810..102T}. A simplified statistical description of clumped winds in HMXBs was developed by \citet{2020A&A...643A...9E}, who assume spherical clumps advected outward from the star.

The dynamical impact of X-ray irradiation on radiatively driven winds in HMXBs has been studied in detail by Krtička and collaborators.
\citet{2012ApJ...757..162K} modeled the wind of the Vela~X-1 supergiant companion including the X-ray irradiation from the NS, showing that photoionization destroys the ions responsible for line driving and produces a photoionized bubble with a stagnating flow around the compact object.
This framework was later generalized by \citet{2015A&A...579A.111K}, who explored the effect of X-ray luminosity and source separation on the winds of massive binary components and identified wind-inhibition regimes in the \(L_{\rm X}\)--optical-depth parameter space.
\citet{2018A&A...620A.150K} further investigated the role of clumping, showing that clumping weakens the effect of X-ray irradiation by favoring recombination and increasing the wind mass-loss rate.

\citet{2009MNRAS.398.2152D} studied the role of structured, clumpy stellar winds in shaping the accretion rate and X-ray variability of supergiant systems, while \citet{2021MNRAS.501.2403B} developed a semi-analytical treatment of wind accretion in eccentric binaries, emphasizing the strong orbital modulation of the accretion flow in systems with large eccentricities.

Emission lines in HMXB winds arise through several channels \citep[e.g.,][]{Kallman_2001a}. Photoexcitation can produce resonance scattering, in which an absorbed continuum photon is re-emitted at the same transition energy but in a different direction. In photoionized gas, recombination (including dielectronic recombination) followed by radiative cascades can also contribute strongly to line formation. In addition, collisional excitation may become important in dense or shock-heated regions \citep[e.g.,][]{2001ApJ...559.1108O}. The relative importance of these processes depends on the local density, temperature, ionization state and radiation field \citep[e.g.,][]{Watanabe_2006}.

The observed emission and absorption features depend on the system geometry and on how the compact object illuminates the wind. As it moves along its orbit, the line of sight and the irradiation pattern change, so the ionization structure and the visibility of line-forming regions vary with orbital phase $\phi$. This phase dependence provides a powerful way to diagnose both the circumstellar environment and the binary configuration \citep[e.g.,][]{Goldstein_2004a,Miskovicova_2016, Aftab_2019,Fogantini_2021,Sanjuro-Ferrin_2021}. However, modeling the ionization structure of the clumpy wind and therefore constraining the origin of the observed emission has proven challenging. Given the recent progress in high resolution X-ray spectroscopy of HMXBs with \textsl{XRISM} (\citealt{2022IJMPD..3130001T}, see e.g. \citealt{Mochizuki_2024} \citealt{Diez_2025} and \citealt{Yamada_2025} for individual sources) and expected order of magnitude improvements with \textsl{NewAthena} \citep{Cruise2025}, we aim to lay the groundwork for such models in this paper.

The stellar wind and its illumination can be described by three sets of parameters. The first set characterizes the orbit (e.g., eccentricity, argument of periastron, and semi-major axis). The second set defines the orientation with respect to the observer (e.g., orbital inclination and orbital phase $\phi$). The third set describes the wind structure and irradiation properties, namely the mass-loss rate ($\dot{M}$), terminal velocity ($v_{\infty}$), wind acceleration parameter ($\beta$), X-ray luminosity ($L_{\mathrm{X}}$), elemental abundances, and, where applicable, the adopted clump expansion prescription. Throughout this paper, we reserve the term wind--radiation parameters for the subset $\{\dot{M},\, v_{\infty},\, L_{\mathrm{X}}\}$, since these are the global quantities that primarily regulate the ionization structure once the orbital parameters and wind structure prescription are fixed. Modeling the wind and its illumination for a given orbital configuration enables predictions and interpretations of the phase-dependent line spectra.

To describe the large-scale wind structure around the donor, we adopt a CAK-like prescription \citep{1975ApJ...195..157C}, characterized by $\dot{M}$ and $v_{\infty}$. The radial velocity profile follows a $\beta$-law; $\beta$ and $v_{\infty}$ are typically constrained from optical and ultraviolet spectroscopy, with $\beta$ in the range $\sim0.5$--2 \citep{2000ARA&A..38..613K}. The wind in the binary therefore exhibits spatial variations in density and ionization that reflect the combined influence of $L_{\mathrm{X}}$ and the local separation between the compact object and the donor.

We employ \texttt{XSTAR}\footnote{\url{https://heasarc.gsfc.nasa.gov/docs/software/xstar/docs/sphinx/xstardoc/docs/build/html/index.html\#}; see also \citet{2001ApJS..133..221K}} to compute charge-state distributions as a function of the ionization parameter. For a tenuous plasma in photoionization equilibrium, ionic fractions are set by the balance between photoionization and recombination, and \texttt{XSTAR} provides these distributions as a function of the ionization parameter $\xi$ defined as $\xi = L_{\rm X}/n\,d_{\rm X}^2$ \citep{1969ApJ...156..943T}, for a given incident spectral energy distribution (SED). It is important to note, however, that the relation between $\log\xi$ and the resulting charge-state distribution is not universal, but depends sensitively on the shape of the ionizing SED.


The compact object acts as a moving X-ray probe of the donor wind: as it orbits, it illuminates and is seen through different regions of the clumpy outflow. This is particularly informative at high inclination, where phase-resolved spectroscopy can exploit (i) periodic orbital modulation driven by geometry (e.g., changes in line-of-sight absorption, occultation, and X-ray shadowing) and (ii) stochastic variability produced by individual clumps crossing the line of sight and/or entering the illuminated region. Together, these signatures can be used to constrain the wind properties and the system geometry.

In this work, we develop a phase-dependent framework that combines a CAK-like wind prescription, clumping, and \texttt{XSTAR} photoionization calculations to map the three-dimensional distribution of ions in HMXBs. The model explicitly includes orbital geometry, occultation effects, and X-ray shadowing, allowing us to identify which ionic regions remain visible at each orbital phase and their relative contribution. We then incorporate these model diagnostics into a Bayesian inference scheme to constrain the global wind--radiation parameters, $\left(\dot{M},\, v_{\infty},\, L_{\mathrm{X}}\right)$, from eclipse spectroscopy.

This paper is structured as follows. Section~2 introduces the two-component stellar-wind parametrization (clumps plus inter-clump components). Section~3 describes the construction of 3D ion maps and the phase-dependent visibility diagnostics. Section~4 outlines the Bayesian inference methodology to constrain the wind--radiation parameters and discusses the limitations of the employed approach. Section~5 applies the workflow to the eclipse observations of XTE~J1855$-$026 and 4U~1700$-$37. We summarize our conclusions in Section~6. 
\section{Stellar-wind parametrization: clumps and inter-clump components}
\label{sec:wind_param}

\subsection{Model}
The stellar wind is not perfectly smooth; instead, overdense structures (clumps) coexist with a tenuous inter-clump medium. In \citet{2020A&A...643A...9E}, the clumpy wind is described in the limiting case, in which all wind mass is concentrated in discrete spherical clumps embedded in a void inter-clump medium. We adopt the same clump geometry and kinematic assumptions (spherical clumps, radial advection, and no merging above an onset radius), but relax the void-interclump approximation by allowing the inter-clump component to carry a finite fraction of the mass-loss rate, hereafter referred to as the inter-clump component.

The large-scale wind stratification is assumed to be stationary and spherically symmetric. The velocity profile follows a standard \(\beta\)-law \cite{1975ApJ...195..157C}.

\begin{align}
\label{eq:cak}
v(r) &= v_\infty \left(1 - \frac{R_\star}{r}\right)^\beta, &
\rho(r) &= \frac{\dot{M}}{4 \pi r^2 v(r)},
\end{align}
where \(v_\infty\) is the terminal velocity, \(R_\star\) the stellar radius, \(\beta\) the wind-acceleration parameter, and \(\dot{M}\) the mass-loss rate. The quantity \(\rho(r)\) denotes the smooth  density.

We describe the stellar wind as a two-component medium of overdense clumps embedded in a tenuous inter-clump component. By mass conservation, the volume-averaged density satisfies
\begin{equation}
\rho(r)= f_V(r)\,\rho_{\rm cl}(r) + \bigl[1-f_V(r)\bigr]\,\rho_{\rm ic}(r),
\label{eq:two_phase_average}
\end{equation}
where \(\rho_{\rm cl}\) and \(\rho_{\rm ic}\) are the clump and inter-clump mass densities, and \(f_V\) is the geometrical clump volume filling factor.

A convenient parametrization is obtained with the dimensionless density ratios
\begin{equation}
f_\rho \equiv \frac{\rho}{\rho_{\rm cl}}, \qquad
\delta_{\rm ic}\equiv \frac{\rho_{\rm ic}}{\rho},
\label{eq:frho_delta_def}
\end{equation}
which generalize the usual volume-filling description to finite inter-clump density. Combining Eqs.~\eqref{eq:two_phase_average} and \eqref{eq:frho_delta_def} gives
\begin{equation}
f_V = \frac{1-\delta_{\rm ic}}{f_\rho^{-1}-\delta_{\rm ic}},
\label{eq:fV_from_frho_delta}
\end{equation}
and the fraction of wind mass carried by clumps is
\begin{equation}
\alpha_{\rm m}\equiv \frac{\dot{M}_{\rm cl}}{\dot{M}}=\frac{f_V\rho_{\rm cl}}{\rho}=\frac{f_V}{f_\rho}.
\label{eq:alpha_mass}
\end{equation}

Clumps are assumed to be injected above an onset radius \(r_0\) and then advected radially with Eq.~\eqref{eq:cak}. If the injection rate is \(\dot{N}\) clumps~s\(^{-1}\), the clump number density (i.e. the number of clumps per volume) follows from continuity:
\begin{equation}
n_{\rm cl}(r)=\frac{\dot{N}}{4\pi r^2 v(r)}.
\label{eq:ncl}
\end{equation}
The expected number of clumps between two radii is therefore
\begin{equation}
N(r_0,r_{\max})=\dot{N}\int_{r_0}^{r_{\max}}\frac{dr}{v(r)}.
\label{eq:Ncl_integral}
\end{equation}

Clumps are taken to be spherical and to carry a fixed mass \(m_{\rm cl}\), while their radius increases with distance from the star. In \citet{2020A&A...643A...9E}, two expansion laws are considered:
\begin{align}
R_{\rm cl}(r) &\propto \left[v(r)\,r^2\right]^{1/3}
\qquad \text{(smooth expansion)}, \label{eq:Rcl_}\\
R_{\rm cl}(r) &\propto r
\qquad \text{(linear expansion)}. \label{eq:Rcl_linear}
\end{align}
For constant \(m_{\rm cl}\), the -expansion case implies \(\rho_{\rm cl}(r)\propto [r^2 v(r)]^{-1}\), that is, the same radial dependence as \(\rho_{\rm ic}(r)\), preserving an approximately constant density contrast across the wind. The linear-expansion case instead produces a steeper radial dilution of clump density. In this work, we follow the example of \citep{2020A&A...643A...9E} and adopt the smooth-expansion law; we refer to their paper for a detailed discussion of the appropriateness of this assumption.

To characterize local irradiation, we use the standard ionization parameter following \citet{1969ApJ...156..943T},
\begin{equation}
\xi = \frac{L_{\rm X}}{n\,d_{\rm X}^2},
\label{eq:xi_def}
\end{equation}
where \(L_{\rm X}\) is the compact-object luminosity, \(d_{\rm X}\) is the distance to the X-ray source, and \(n=\rho/(\mu m_{\rm p})\) is the local particle number density, with \(\mu\) the mean molecular weight and \(m_{\rm p}\) the proton mass. We further include X-ray shadowing by the donor star, but, in this first paper, not the absorption and emission effects in the wind itself.

In this first-order analysis, we neglect transmission and absorption effects, as our primary objective is to identify the regions where each ion is expected to be most abundant mapping local ionic fractions. This constitutes a computationally inexpensive strategy for obtaining the most informative results from an initial, approximate calculation and for demonstrating that such a first-order approach is capable of providing valuable information, both theoretical and experimental, on HMXB systems and their clumping prescriptions.

\subsection{Example}
\label{sec:example}

As an illustrative  example, we adopt representative HMXB parameters inspired by XTE~J1855-026 \citep{2022MNRAS.512..304S}. The adopted system parameters are listed in the second column of Table~\ref{tab:system_overview}. We adopted a simple \texttt{powerlaw} spectral energy distribution (SED) with a photon index of 1.2 to represent the direct emission from the neutron star, primarily for reproducibility. We stress, however, that the predicted charge-state-distribution depends critically on the assumed SED and the example presented here just serves as a case study for typical neutron star HMXB parameters. We note that quantitative constraints on the ionization parameter for a given source will generally require a recalculation of the ion maps using a photoionization code (e.g., \texttt{XSTAR) } with an SED appropriate for that specific source.  
For the clumping setup, we assume that 98\% of the wind mass is contained in clumps ($\alpha_m \simeq 0.98$). This choice is motivated by the $\xi$ contrast between the two ionization plasmas identified in the eclipse spectral analysis, which we associate with the clump and inter-clump components of the wind under the clumping prescription described in Sect.~\ref{sec:wind_param}. This point is further discussed in Sect.~\ref{sec:application}. We also adopt a characteristic clump size of \(0.015\,R_\star\) at a reference radius of \(2\,R_\star\), an injection above \(1.2\,R_\star\), and a density contrast \(\rho_{\rm cl}/\rho_{\rm ic}=1000\). The corresponding clumping parameters are summarized in Table~\ref{tab:clumping_summary}.

\begin{table}[hb]
\centering
\caption{Clumping parameters adopted in the example model.}
\label{tab:clumping_summary}
\begin{tabular}{lc}
\hline
Parameter & Value \\
\hline
\(f_\rho \equiv \rho/\rho_{\rm cl}\) & \(4.77\times10^{-2}\) \\
\(\rho_{\rm cl}/\rho\) & \(21.0\) \\
\(\delta_{\rm ic} \equiv \rho_{\rm ic}/\rho\) & \(2.10\times10^{-2}\) \\
\(f_V\) & \(4.67\times10^{-2}\) \\
\(\alpha_{\rm m}\) & \(0.98\) \\
\(\rho_{\rm cl}/\rho_{\rm ic}\) & \(1.00\times10^{3}\) \\
\(\dot{N}\) & \(1.84~\mathrm{s^{-1}}\) \\
\(m_{\rm cl}\) & \(6.69\times10^{19}~\mathrm{g}\) \\
\(N_{\rm snapshot}\) & \(\approx 8.84\times10^{5}\) \\
\(\dot{M}_{\rm cl}\) & \(1.96\times10^{-6}~M_\odot~\mathrm{yr^{-1}}\) \\
\(\dot{M}_{\rm ic}\) & \(4.00\times10^{-8}~M_\odot~\mathrm{yr^{-1}}\) \\
\hline
\end{tabular}
\end{table}

\begin{figure*}[h]
\centering
\includegraphics[width=1\textwidth]{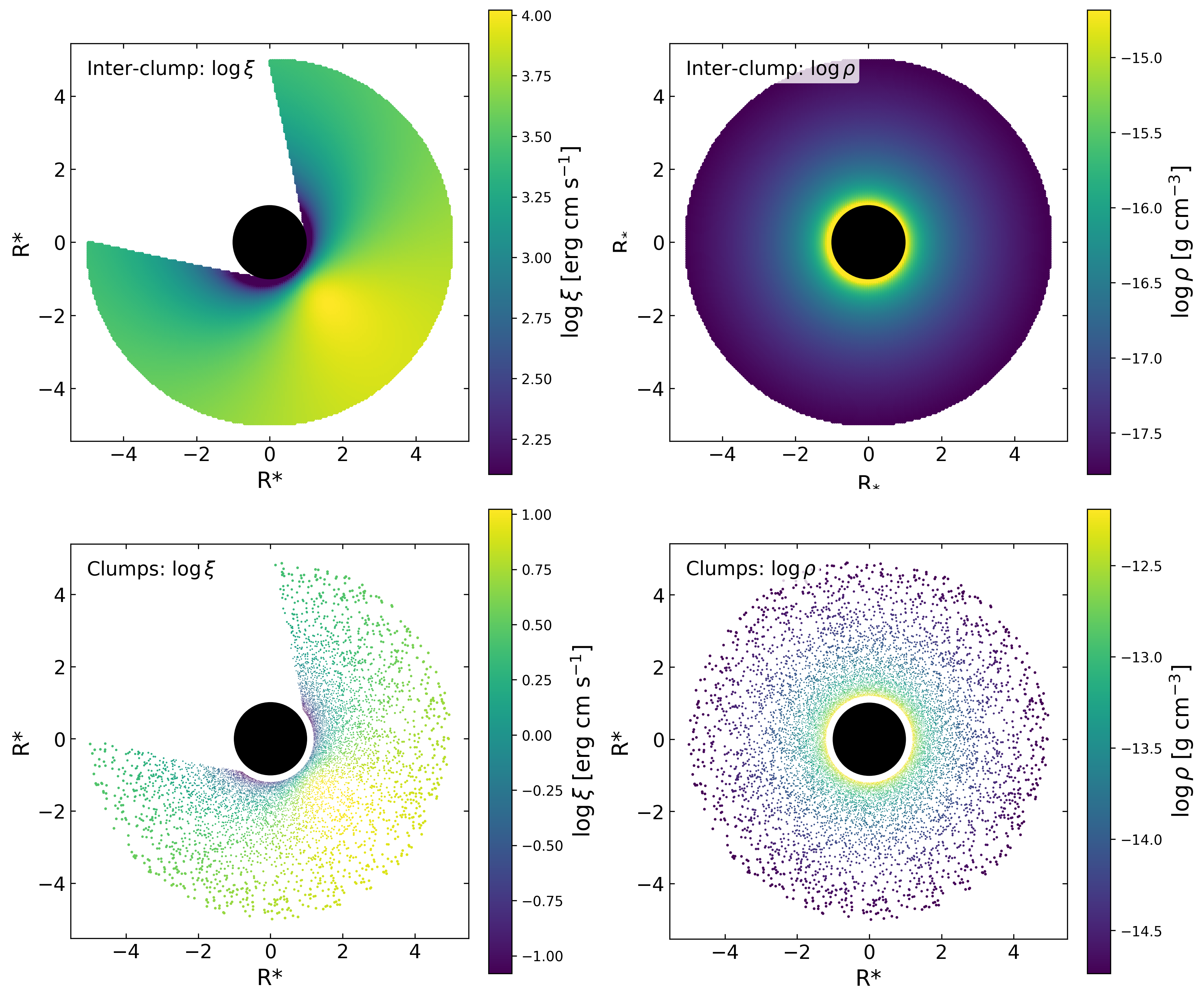}
\caption{Example 2D density and ionization maps of the inter-clump and clump components of the wind computed with the parameters in Tables~\ref{tab:system_overview} and \ref{tab:clumping_summary}. From left to right, the panels show $\log_{10}\rho$ in the inter-clump medium, $\log_{10}\xi$ in the inter-clump medium, $\log_{10}\rho$ in the clumps, and $\log_{10}\xi$ in the clumps. Each panel uses an independent colour normalisation to enhance internal structure, and the clump panels display a random subset for clarity.}
\label{fig:chi_rho_example}
\end{figure*}

In Fig.~\ref{fig:chi_rho_example}, the upper row shows the inter-clump \(\xi\) and \(\rho\) 2D maps, whereas the bottom row panels show the corresponding clump \(\xi\) and \(\rho\) maps. In this configuration, \(\sim 98\%\) of the wind mass is concentrated within only \(\sim 5\%\) of the total volume. The maps show that the inter-clump medium reaches substantially higher ionization levels than the clumps, because of its lower density.

\section{Ion maps and phase-dependent visibility}

We use the example introduced in Sect.~\ref{sec:example} to construct three-dimensional distributions of ionic number density (hereafter, ion maps) and to examine their phase-dependent visibility. These maps are used to identify the regions of the two-component (clump + inter-clump) wind where a given ion is expected to be most abundant.

\begin{figure}[ht]
\centering
\includegraphics[trim={0cm 0cm 0cm 0cm},width=1\columnwidth]{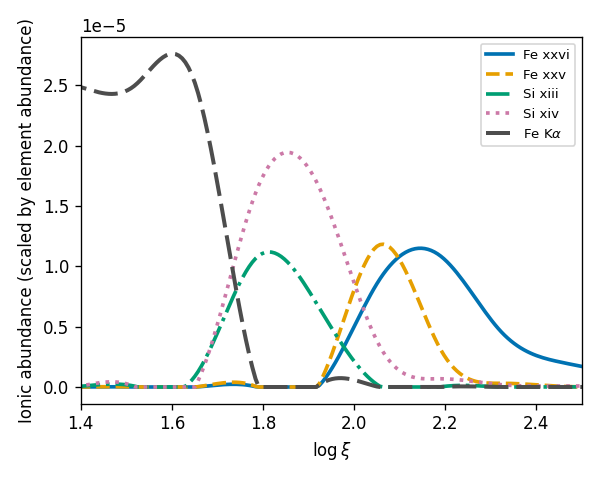}
\caption{Ionic fraction as a function of $\log \xi$ for Fe\,XXVI, Fe\,XXV, Si\,XIII, Si\,XIV, and Fe K$\alpha$ emission, calculated for the spectral energy distribution described in Sect.~\ref{sec:example}.}
\label{fig:ion_abund}
\end{figure}

For each grid cell (and for each clump), we derive the hydrogen number density, $n_{\rm H}$, from the local mass density, $\rho$, and adopt elemental abundances $A_{\rm el}\equiv n({\rm el})/n({\rm H})$ from the selected \texttt{xdef} \texttt{XSTAR} abundance table, assuming a tenuous plasma in photoionization equilibrium. For illustration, the ionic fraction as a function of $\log\xi$ is shown in Fig.~\ref{fig:ion_abund}. Combining these abundances with the ionic fractions $f_{{\rm el},q}$ from the local charge-state distribution, the number density of a given ion is

\begin{equation}
n_{\rm ion}=n_{\rm H}\,A_{\rm el}\,f_{{\rm el},q}.
\end{equation}

\subsection{Effects of binary system geometry}

The ion maps allow us to quantify, as a function of orbital phase, the fraction of the ionic volume that remains observable, i.e. neither eclipsed by the donor nor located in the X-ray shadow. This observable fraction varies because different regions of the wind are alternately illuminated and occulted as the neutron star moves along its orbit. For circular systems with low or moderate inclination, the modulation is expected to be weak in an approximately axisymmetric geometry; any residual variability mainly reflects the stochastic distribution of clumps. At higher inclinations and/or non-zero eccentricity, the modulation generally increases owing to phase-dependent occultation and changes in the line-of-sight column density. In eccentric systems, additional variability may arise from the phase-dependent orbital separation, which alters the local density around the compact object and modifies the ionization balance.

\begin{figure*}[ht]
\centering
\includegraphics[trim={0cm 0cm 0cm 3cm},width=1\textwidth]{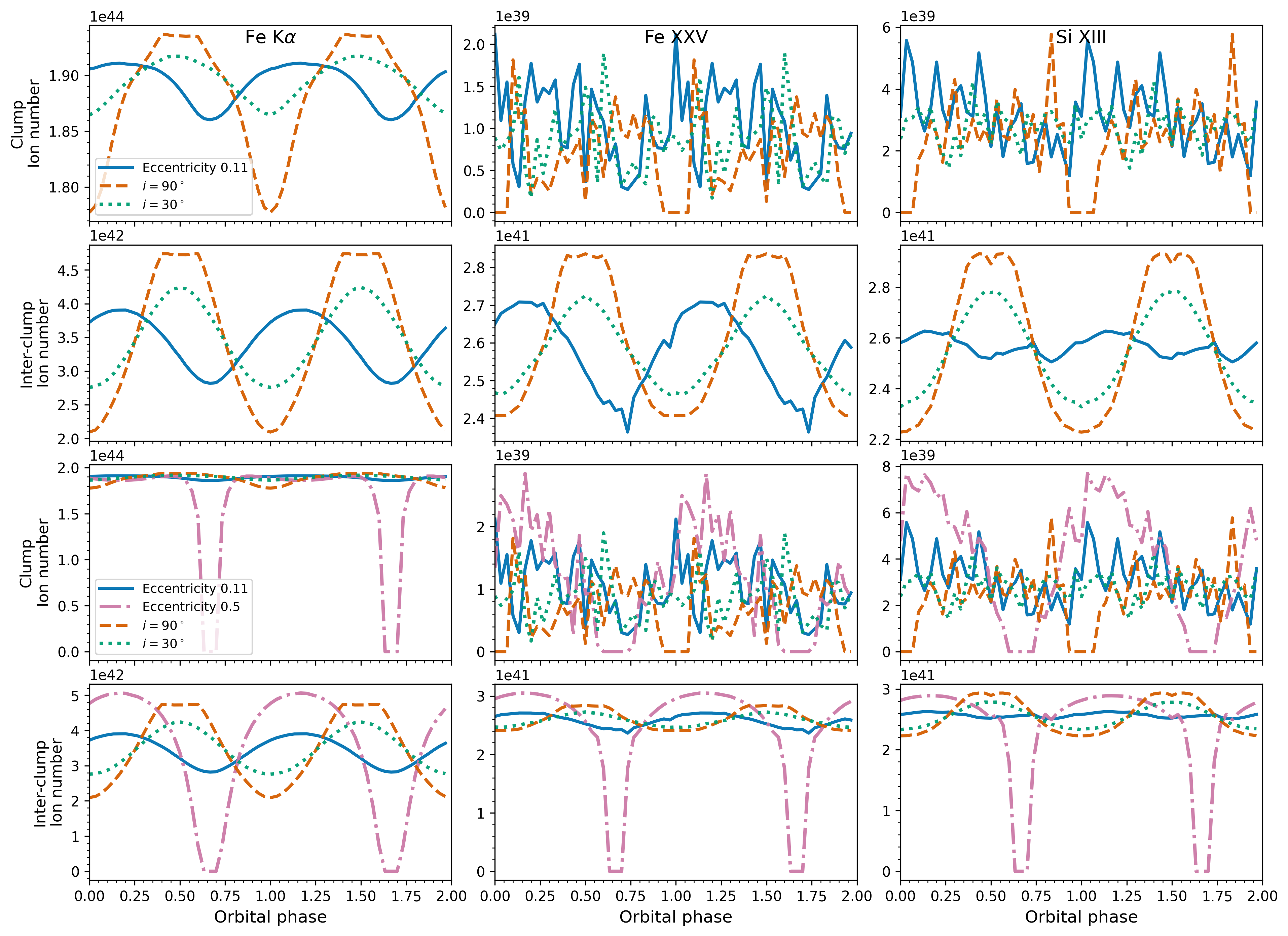}
\caption{
Relative observable ionic volume as a function of orbital phase.
Columns, from left to right, correspond to Fe K$\alpha$, \ion{Fe}{xxv}, and \ion{Si}{xiii}.
The first and second rows show the clump and inter-clump contributions, respectively, 
for three reference orbital configurations: 
a circular edge-on case (\(i=90^\circ,\,e=0\)), 
a circular moderate-inclination case (\(i=30^\circ,\,e=0\)), 
and a mildly eccentric case (\(e=0.11\)).
The third and fourth rows show the same quantities after including an additional 
high-eccentricity configuration (\(e=0.50\)), to highlight the enhanced orbital 
variability produced by a more eccentric orbit.
In each panel, the green dotted line indicates the circular 
\(i=30^\circ,\,e=0\) case; the orange dashed line indicates the circular 
\(i=90^\circ,\,e=0\) case; the blue solid line indicates the mildly eccentric 
case with \(e=0.11\); and, in the two lower rows, the magenta dash-dotted line 
indicates the highly eccentric case with \(e=0.50\).}
\label{fig:line_orbital_ev}
\end{figure*}

To illustrate these effects, we consider four representative orbital configurations:
(i) \(i=90^\circ,\,e=0\),
(ii) \(i=30^\circ,\,e=0\),
(iii) \(i=0^\circ,\,e=0.11\), and
(iv) \(i=0^\circ,\,e=0.50\).
The corresponding orbital evolution of the observable ionic volume is shown in Fig.~\ref{fig:line_orbital_ev} for the clump and inter-clump components of the wind.

We define the Fe K$\alpha$ emission from the XSTAR transitions corresponding to Fe 2p→1s lines between 6.38 and 6.42 keV. In order to ensure that only the cold, weakly ionized Fe K$\alpha$ component is included, we limit the selected ions to \ion{Fe}{i}--\ion{Fe}{xv} XV. This selection is intended to isolate the low-ionization Fe species that are expected to produce the observed fluorescent Fe K$\alpha$ emission.

In this particular example, Fe K$\alpha$ emission is associated mainly with the illuminated donor-facing wind and is typically not fully eclipsed. In contrast, \ion{Fe}{xxv} and \ion{Si}{xiii} are concentrated predominantly in highly ionized regions near the neutron star and can be completely eclipsed during parts of the orbit.

In the clumpy component, \ion{Fe}{xxv} appears only when dense clumps are located very close to the NS; accordingly, the clumpy \ion{Fe}{xxv} contribution in Fig.~\ref{fig:line_orbital_ev} shows strong stochastic variability. The moderate-inclination case remains close to the maximum observable value for highly ionized species such as \ion{Fe}{xxv}, because most of the corresponding ionic volume remains visible throughout the orbit. Fe K$\alpha$ emission, however, traces a more extended region and is therefore less sensitive to complete eclipse.

The case of highly eccentric wind-fed systems has been studied in detail by  \citet{2021MNRAS.501.2403B}, who presented a semi-analytical treatment of  wind accretion onto a neutron star in supergiant HMXBs with eccentric orbits. Their model couples the orbital modulation of the mass-accretion rate with the X-ray photoionization of the stellar wind, showing that eccentricity can produce  strong phase-dependent changes in the relative velocity, accretion rate, and X-ray luminosity. 
Our results are qualitatively consistent with this picture. Although the present  work does not attempt to model the accretion luminosity itself, increasing the eccentricity produces the strongest orbital modulation of the observable ionic volume. This is illustrated in Fig.~\ref{fig:line_orbital_ev}, where the two  lower rows show that the highly eccentric orbital configurations lead to much larger variations of the observed ionic volume along the orbit than the nearly circular cases, even with the highest inclination. 
This behavior is expected, since higher eccentricity implies larger changes in  the neutron star separation from the donor, and therefore in the local wind density, ionization balance, and illuminated fraction of the wind.

We emphasize that these maps trace ion-location volume, not emergent line flux. Direct comparison with observed line fluxes requires radiative transfer effects, including absorption and transmission, to be taken into account.

\begin{figure*}[ht]
\centering
\includegraphics[trim={0cm 0cm 0cm 0cm},width=1\textwidth]{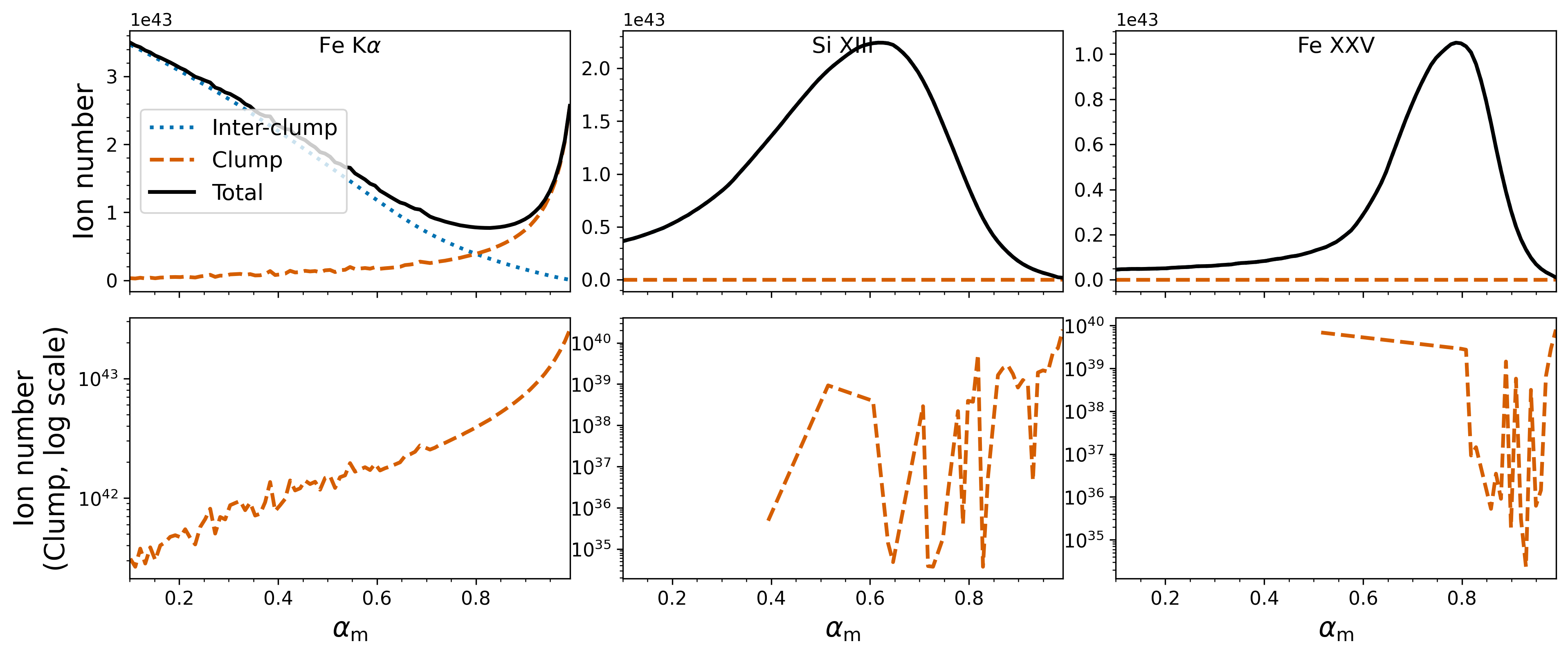}
\caption{
Upper row: observable ionic volume as a function of \(\alpha_{\rm m}\), the wind mass fraction in clumps, at fixed orbital phase \(\phi=0.25\). Curves are shown separately for inter-clump, clump and total contributions for each representative ion. Lower panel: Only clump contribution in logarithmic scale.
}
\label{fig:ion_clump_ev}
\end{figure*}

\subsection{Effects of the clumping level}

In addition to geometry, the clumping level also affects the ionization structure and therefore the observable ionic volumes (Fig.~\ref{fig:ion_clump_ev}). For a fixed neutron star orbital phase (\(\phi=0.25\) in this example), we varied \(\alpha_{\rm m}\), the fraction of wind mass contained in clumps, and quantified the corresponding change in observable volume for representative ions.

As \(\alpha_{\rm m}\) increases, a larger fraction of the wind mass is transferred to the dense clump phase, while the inter-clump medium becomes comparatively more tenuous. This increases the contrast between the two phases in both density and ionization state, and modifies the relative observable ionic volumes traced by different species.

In this configuration, Fe K\(\alpha\) becomes clump-dominated for \(\alpha_{\rm m}\gtrsim 0.85\), i.e. when more than \(\sim 75\%\) of the wind mass is locked into clumps. This reflects its preferential association with denser and less ionized material. More highly ionized lines, by contrast, remain predominantly associated with the inter-clump component, which continues to trace the more diluted and strongly irradiated phase of the wind.

For the highly ionized species, however, the behavior reflects an interplay between ionization state and mass distribution. When \(\alpha_{\rm m}\) approaches unity, the larger number of clumps increases the probability that some of them lie sufficiently close to the neutron star to become strongly ionized, at the same time, the inter-clump medium is more tenuous and exhibits a higher $\xi$. As a result, the clump contribution to highly ionized species can increase again at large \(\alpha_{\rm m}\), suggesting the presence of a transition or mild saturation regime. This non-monotonic trend is seen for \ion{Si}{xiii} and \ion{Fe}{xxv} for example. In the inter-clump component, its observable volume increases as the medium becomes more tenuous and reaches a more favorable ionization state. At still larger \(\alpha_{\rm m}\), however, the ionization parameter becomes too high for \ion{Si}{xiii} to dominate, and higher-ionization species such as \ion{Fe}{xxv} become favored instead. This suggests that, for \(\alpha_{\rm m}\) close to unity, the inter-clump medium reaches characteristic values around \(\log\xi\sim 1.5-2.5\) in this particular configuration.

This behavior suggests that the relative prominence of low- and high-ionization tracers may provide a useful diagnostic of the clumping level. The transition value found here should, however, be regarded as configuration-dependent, since the balance between clump and inter-clump contributions also depends on orbital phase, visibility, and the adopted wind structure.

\section{Inference of wind--radiation parameters}
\label{sec:wind_inference}

Our main goal is to constrain the global wind--radiation regime implied by the ionization state inferred from eclipse spectra. To this end, we adopt an observational estimate of the ionization parameter, derived either from spectral modeling or from diagnostic emission lines, and use it to constrain the wind--radiation parameters \(L_{\mathrm X}\), \(v_{\infty}\), and \(\dot M\) within a Bayesian framework.

Eclipse spectra are particularly informative for this problem. During eclipse, the donor star occults the direct neutron star continuum as well as most of the innermost, highest-ionization region, so the maximum observable ionization remains finite and physically interpretable. In addition, eclipse spectra often show enhanced line-to-continuum ratios \citep[e.g.][]{2002ApJ...564L..21S, 2015ApJ...810..102T,Aftab_2019,2021MNRAS.501.5646M}, which improves line identification and facilitates ionization diagnostics.

For a fixed wind structure, orbital configuration, and visibility mask, the ionization parameter can be written as
\begin{equation}
\label{eq:xi_profile_new}
\xi(d,r)=\xi_{0}\,\chi(d,r),
\end{equation}
where
\begin{equation}
\label{eq:xi0_new}
\xi_{0}=
\frac{L_{\mathrm X}\,\mu m_{\mathrm p}\,v_{\infty}}{\dot M},
\end{equation}
and \(\chi(d,r)\) is a dimensionless geometric--kinematic factor. For the clump and inter-clump components, respectively, we write
\begin{equation}
\label{eq:geom_fac_cl}
\chi_{\rm cl}(d,r)=
\frac{f_V(r)}{\alpha_m(r)}
\left(1-\frac{1}{d}\right)^{\beta}
\left(\frac{d}{r}\right)^2 ,
\end{equation}
\begin{equation}
\label{eq:geom_fac_sm}
\chi_{\rm ic}(d,r)=
\frac{1-f_V(r)}{1-\alpha_m(r)}
\left(1-\frac{1}{d}\right)^{\beta}
\left(\frac{d}{r}\right)^2 .
\end{equation}
Here, \(d\) denotes the distance to the donor center and \(r\) the distance to the X-ray source, both expressed in units of \(R_\star\).
Similarly, in \cite{2012MNRAS.421.2820O}, employing an analogous theoretical framework, the authors demonstrate that clumping in the stellar wind leads to a reduction of the ionization parameter by a factor of $f_V(r)$.

Once geometry and visibility are fixed, the dominant global dependence of the ionization state is captured by
\begin{equation}
\xi_0 \propto \frac{L_{\mathrm X} v_\infty}{\dot M}.
\end{equation}
Both wind phases share this same global scaling, while local differences are encoded in \(\chi\) and in the corresponding statistics computed over the visible wind volume. See Sec. \ref{app:calc_xi} for detailed calculations.

A natural framework for this inference is Bayesian analysis \citep{2010arXiv1002.2080R}, which combines prior information with observational constraints. By Bayes' theorem,
\begin{equation}
P(\theta \mid D) \propto P(D \mid \theta)\,P(\theta),
\end{equation}
where \(P(\theta \mid D)\) is the posterior distribution of the parameter vector \(\theta\), \(P(\theta)\) is the prior, and \(P(D \mid \theta)\) is the likelihood of the data \(D\) given \(\theta\).

In all inference runs presented here, we fix an intrinsic log-space scatter term of \(\sigma_{\mathrm{dex}}=0.1\). This quantity is not sampled as a free parameter, but is instead adopted as a constant nuisance term to account for residual model--data dispersion not captured by the deterministic wind prescription, such as small-scale variability, calibration systematics, or simplified physics. In logarithmic terms, this corresponds to a characteristic scatter of \(\pm 0.1\) dex around the model prediction, that is, a multiplicative uncertainty of about a factor of \(10^{0.1}\approx1.26\) in linear space. In practice, if the model predicts \(\log \xi = 5\), values in the range \(4.9 \lesssim \log \xi \lesssim 5.1\) are treated as consistent within this intrinsic scatter.

We adopt wide uniform priors spanning the physically plausible parameter ranges established for each source (Table~\ref{tab:system_overview}). Posterior inference is performed using the No-U-Turn Sampler (NUTS), with four independent Markov chains, \(10^{3}\) warm-up iterations, and \(10^{4}\) post-warm-up draws per chain. Convergence is assessed using the rank-normalized split-\(\hat{R}\) statistic and effective sample sizes, following \citet{10.1214/20-BA1221}. Within this framework, we consider three representative formulations of the inference problem.

In summary, we explicitly compute and fix both \(\chi_{\rm cl}(d,r)\) and \(\chi_{\rm ic}(d,r)\). To this end, we adopt fixed orbital parameters, system orientation with respect to the observer, and a prescribed value of the \(\beta\) exponent. For a given configuration, the observed \(\xi\) then depends on the stellar wind properties and radiative parameters. The Bayesian inference procedure systematically explores the physically admissible region of parameter space and identifies those parameter combinations that most accurately reproduce the \(\log \xi\) values derived from the spectral analysis. The resulting posterior distributions thus delineate the parameter regimes most strongly favored by the data, while the use of four independent Markov chains enables a direct evaluation of convergence and the robustness of the inferred parameters.

Inference of the clumping prescription, orbital parameters, and viewing geometry is in principle also possible, but would be considerably more expensive computationally. The factorized form of $\xi$ provides an efficient first-order approach for extracting the main physical information.

\subsection{Single-constraint case: maximum observable ionization}
\label{sec:Single-constraint case: maximum observable ionization}

A first observable is the maximum ionization reached within the visible wind. Rather than using the absolute maximum, we adopt the mass-weighted 95th percentile of the visible \(\xi\) distribution in order to reduce sensitivity to extreme values and to ionization states that contribute negligibly to the total wind mass. The forward model is then
\begin{equation}
\log_{10}\xi_{\max}^{\mathrm{mod}}
=
\log_{10}\xi_0 + C_{\max},
\end{equation}
where \(C_{\max}\) is a geometry-dependent constant obtained from the precomputed \(\chi(d,r)\) map for the adopted orbital configuration and clump setup. Given an observed value \(\log_{10}\xi_{\max}^{\mathrm{obs}}\), we assume
\begin{equation}
\log_{10}\xi_{\max}^{\mathrm{obs}}
\sim
\mathcal N\!\left(\log_{10}\xi_{\max}^{\mathrm{mod}},\,\sigma_{\mathrm{dex}}^2\right).
\end{equation}

\begin{figure*}[ht]
\centering
\includegraphics[trim={8cm 0cm 8cm 0cm}, width=1\textwidth]{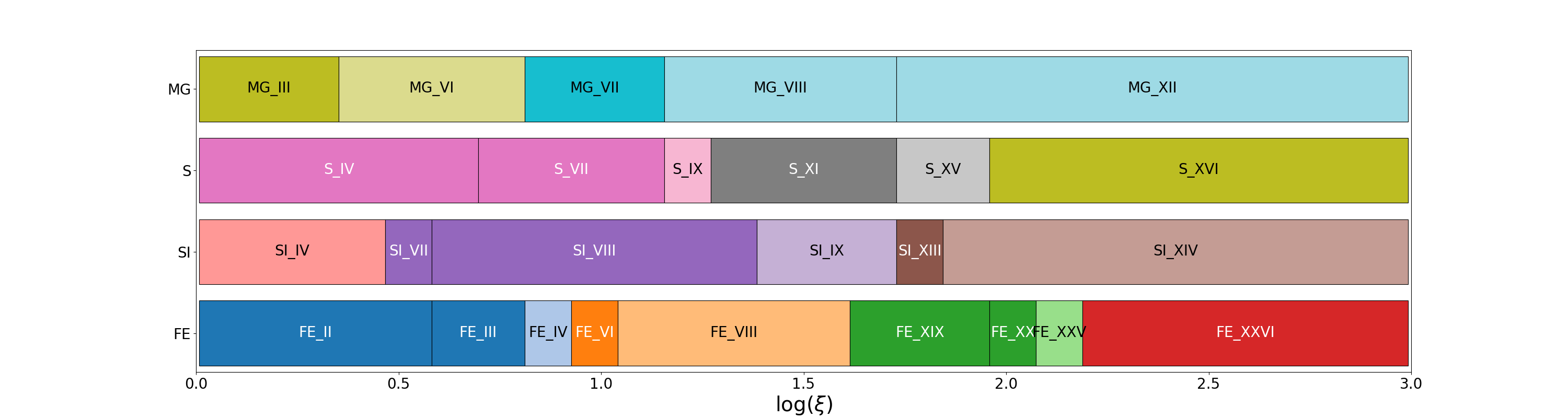}
\caption{Most prominent ion per element as a function of the ionization parameter \(\xi\) for the direct neutron star emission SED reported for the XTE~J1855$-$026 out of eclipse spectra in \cite{2018JApA...39....7D}.}
\label{fig:mainionmatrix}
\end{figure*}

Within the optically thin, photoionized-plasma framework, with \texttt{XSTAR} for a given SED we construct an ion matrix that identifies, for each chemical element, the dominant ion as a function of \(\xi\) (Fig.~\ref{fig:mainionmatrix}). The set of ionic species detected in an observed X-ray spectrum constrains the range of ionization parameters present in the visible plasma, and in particular the upper ionization level sampled during eclipse. In this case, as the SED we took a \texttt{powerlaw} of photon index 1.2.

\subsection{Two-component case: joint clump/inter-clump constraints}
\label{sec:Two-component case: joint clump/inter-clump constraints}

When the spectral fit requires two photoionized components (e.g. two \texttt{photemis} zones), we use two observational ionization constraints, \(\xi_{\mathrm{cl}}^{\mathrm{obs}}\) and \(\xi_{\mathrm{ic}}^{\mathrm{obs}}\), associated predominantly with the clump and inter-clump phases, respectively. The target values are taken from \texttt{photemis} (or an equivalent self-consistent photoionization model); \texttt{photemis} is the XSTAR-based additive emission model, consistent with the atomic and ionization framework of \texttt{warmabs} \citep{Kallman_2001a}.

For each wind phase \(k\in\{\mathrm{cl},\mathrm{ic}\}\), we compute from the visible 3D map a summary constant \(C_k\), defined from the chosen weighted summary of the corresponding \(\log_{10}\chi\) distribution. In the applications shown below, we use the particle-weighted mean. The forward model is
\begin{equation}
\mu_k=\log_{10}\xi_0 + C_k,
\end{equation}
where \(\mu_k\) is the modeled mean log-ionization for phase \(k\). The two constraints are then fitted simultaneously through the joint likelihood
\begin{align}
\log_{10}\xi_{\mathrm{cl}}^{\mathrm{obs}} &\sim
\mathcal N\!\left(\mu_{\mathrm{cl}},\sigma_{\mathrm{dex}}^2\right),\\
\log_{10}\xi_{\mathrm{ic}}^{\mathrm{obs}} &\sim
\mathcal N\!\left(\mu_{\mathrm{ic}},\sigma_{\mathrm{dex}}^2\right).
\end{align}

This is the most direct way to combine both wind phases within a single inference problem, under the assumption that they are governed by the same global wind--radiation parameters but sample different regions of the fixed ionization structure.

\subsection{Separated-component case}
\label{sec:Separated-component case}

As a complementary test, the two wind phases can also be fitted independently. In this case, we perform two separate inferences, one using the clump-associated ionization constraint and one using the inter-clump constraint. For each phase \(k\in\{\mathrm{cl},\mathrm{ic}\}\), the forward model is
\begin{equation}
\log_{10}\xi_{k}^{\mathrm{mod}}
=
\log_{10}\xi_0 + C_{k},
\end{equation}
where \(C_k\) is the phase-dependent summary constant computed from the corresponding precomputed \(\chi(d,r)\) map, visibility mask, and phase selection. Given the observed ionization \(\log_{10}\xi_{k}^{\mathrm{obs}}\), we assume
\begin{equation}
\log_{10}\xi_{k}^{\mathrm{obs}}
\sim
\mathcal N\!\left(\log_{10}\xi_{k}^{\mathrm{mod}},\,\sigma_{\mathrm{dex}}^2\right).
\end{equation}

Unlike the joint two-component case, this formulation does not force both phases to be explained simultaneously by a single posterior. Instead, it provides two independent effective inferences that can be compared as a consistency check and as a way to assess how differently the two phases sample the fixed wind structure.

\subsection{Model limitations}
\label{sec:model_limitations}

Before applying the framework to two representative HMXB eclipse observations, we emphasize an important methodological caveat.

The proposed workflow constrains the parameters entering $\xi_{0}$ for a fixed structural term, $\chi(d,r)$ (Eqs.~\ref{eq:xi_profile_new} and \ref{eq:xi0_new}), for a given system configuration and orbital phase.

At first order, the ionization-driven observable is expected to depend primarily on the composite quantity
\begin{equation}
\eta \equiv \log_{10}\!\left(\frac{L_X\,v_\infty}{\dot{M}}\right),
\end{equation}
rather than on each parameter independently. As an example, we performed boundary diagnostics on the posterior samples using wide, non-informative but physically motivated prior domains.:
\begin{equation}
\begin{aligned}
L_X &\in [10^{33},10^{37}]~\mathrm{erg\,s^{-1}},\\
\dot{M} &\in [10^{-8},10^{-5}]~M_\odot\,\mathrm{yr^{-1}},\\
v_\infty &\in [100,4000]~\mathrm{km\,s^{-1}}.
\end{aligned}
\end{equation}

These intervals were chosen to keep solutions within plausible wind-fed HMXB regimes while allowing for a broad parameter exploration. Using the configuration from the XTE~J1855$-$026 example (see Sec. \ref{sec:example}) we observed that the posterior exhibits strong accumulation at prior boundaries:
\begin{itemize}
    \item $L_X$: $\sim 99\%$ of samples near the upper bound,
    \item $\dot{M}$: $\sim 100\%$ of samples near the lower bound,
    \item $v_\infty$: $\sim 44\%$ of samples near the upper bound.
\end{itemize}
This indicates that individual constraints are boundary-limited in the current setup. Hence, although the priors are astrophysically motivated and regularize the inverse problem, they do not fully remove intrinsic degeneracy. Individual estimates of $(L_X,\dot{M},v_\infty)$ should therefore be interpreted as prior-conditional, rather than purely data-driven. The most robustly constrained information is the preferred direction in parameter space (high $L_X$, low $\dot{M}$, and frequently high $v_\infty$). This degeneracy can be partially regularized by adopting system-specific, physically motivated parameter bounds. Therefore, parameter-level constraints should be interpreted within the current ion-location framework and fixed structural assumptions.

\section{Application to XTE~J1855$-$026
and 4U~1700$-$37}
\label{sec:application}

This work is a methodological study in which we use, as illustrative examples, two previously analyzed eclipse observations of HMXBs \citep{10.1093/mnras/staa3956,2022MNRAS.512..304S}. These systems are particularly suitable because both observations were modeled with \texttt{photemis}, explicitly accounting for the $\xi$ clump and inter-clump contributions to the observed emission. 

4U~1700$-$37, discovered with \textit{Uhuru} \citep{1973ApJ...181L..43J}, is an HMXB comprising the O6\,Ia supergiant HD\,153919 (V884\,Sco) and a compact companion. The donor is the earliest-type star currently identified in a Galactic HMXB. The compact object orbits deep within the inner wind, and its persistent X-ray emission strongly ionizes the outflow. Although most system parameters are relatively well constrained, the nature of the compact object remains uncertain because no coherent pulsations have been detected at X-ray or other wavelengths. The 2--200~keV spectrum reported by \citet{1999A&A...349..873R} differs from that of black-hole candidates such as Cyg~X-1 and is instead qualitatively more consistent with accreting neutron-star systems. Eclipse spectra of 4U~1700$-$37 were analyzed in detail by \citet{10.1093/mnras/staa3956}.

XTE~J1855$-$026 was discovered with the \emph{Rossi X-ray Timing Explorer} (\emph{RXTE}) \citep{Corbet_1999}. The system hosts a neutron star with X-ray pulsations at $\simeq 361$~s and an orbital period of $\sim 6$~d. Eclipse-duration analysis by \citet{Corbet_2002} implies a massive companion with a radius consistent with a B0\,I supergiant donor. This classification is supported by spectral energy distribution (SED) modeling \citep{2013ApJ...764..185C} and refined by optical spectroscopy, which constrains the donor to a B0\,Iaep spectral type \citep{gonzalezgalan2016fundamental}.

\begin{table}
\caption{System overview for 4U~1700$-$37 and XTE~J1855$-$026: observation log and adopted parameters.}
\label{tab:system_overview}
\centering

\setlength{\tabcolsep}{5pt}
\begin{tabular}{lcc}
\hline
Parameter & 4U~1700$-$37 & XTE~J1855$-$026 \\
\hline
\multicolumn{3}{c}{System} \\
\hline
\multicolumn{3}{l}{Stellar parameters} \\
Donor spectral type & O6\,Iafpe\tablefootmark{c} & B0\,Iaep\tablefootmark{d,e} \\
$R_{\star}$ [$R_\odot$] & $22\pm2$\tablefootmark{f} & $22\pm2$\tablefootmark{f} \\
$M_{\star,1}$ [$M_\odot$] & $46\pm5$\tablefootmark{f} & $21\pm2$\tablefootmark{f} \\
$M_{\star,2}$ [$M_\odot$] & $1.96\pm0.19$\tablefootmark{f} & $1.41\pm0.24$\tablefootmark{f} \\
\hline
\multicolumn{3}{l}{Wind driving} \\
$\dot M$ [$M_\odot\,\mathrm{yr}^{-1}$] &
$2.5^{+1.5}_{-1.3}\times10^{-6}$\tablefootmark{c} &
$<(0.2$--$1.1)\times10^{-5}$\tablefootmark{f} \\
$\beta$ & 1.2 & 1.5 \\
$v_{\infty}$ [km\,s$^{-1}$] & $1850\pm550$\tablefootmark{c,f} & $620\pm190$\tablefootmark{f} \\
$L_{\mathrm X}$ [erg\,s$^{-1}$] & $3.0\times10^{35}$\tablefootmark{a} & $1.05\times10^{36}$\tablefootmark{b} \\
\hline
\multicolumn{3}{l}{Orbital parameters} \\
Orbital period [d] & 3.411581(7)\tablefootmark{f} & 6.07415(8)\tablefootmark{f} \\
$i$ [deg] & $62\pm2$\tablefootmark{f} & $71\pm2$\tablefootmark{f} \\
$a/R_{\star}$ & $1.6^{+0.5}_{-0.4}$\tablefootmark{c} & 1.8\tablefootmark{b} \\
Orbital phase & 0.848--0.044 & 0.99--0.11 \\
\hline
\multicolumn{3}{c}{$\xi$ observed and clumping prescription} \\
\hline
\multicolumn{3}{l}{Observed $\xi$} \\
$\log_{10}\xi_{\mathrm{cold}}$ & $\sim -1.0$\tablefootmark{a} & 0.36\tablefootmark{b} \\
$\log_{10}\xi_{\mathrm{hot}}$ & $\sim 2.38$\tablefootmark{a} & 3.7\tablefootmark{b} \\
$\log_{10}\xi_{\mathrm{max}}$ & $\sim 2 $\tablefootmark{a} & $\sim 2.25$\tablefootmark{b} \\
\hline
\multicolumn{3}{l}{Clumping prescription from EM ratios \tablefootmark{a,b}} \\
$\alpha_m$ & $\sim 0.95$ & $\sim 0.90$ \\
$D_{\mathrm{cl/sm}}$ & $\sim 300$ & $\sim 180$ \\
\hline
\multicolumn{3}{l}{Clumping prescription from $\xi$ ratios (see Eq.~\ref{eq:ion_ratios})\tablefootmark{a,b}} \\
$\alpha_m$ & $\sim 0.99$ & $\sim 0.98$ \\
$D_{\mathrm{cl/sm}}$ & $\sim 1900$ & $\sim 1000$ \\
\hline
\end{tabular}
\tablefoot{
\tablefoottext{a}{\citet{10.1093/mnras/staa3956}.}
\tablefoottext{b}{\citet{2022MNRAS.512..304S}.}
\tablefoottext{c}{\citet{2020A&A...634A..49H}.}
\tablefoottext{d}{\citet{Corbet_1999}.}
\tablefoottext{e}{\citet{gonzalezgalan2016fundamental}.}
\tablefoottext{f}{\citet{falanga}.}
}
\end{table}

\subsection{The system settings: Clumping prescription and  orbital parameters}
To initiate the analysis, we first adopted a specific clumping prescription. Under the assumption that clumps are isotropically distributed throughout the circumstellar wind, and therefore share the same $\xi_{0}$ and follow the same $\beta$-velocity law as the inter-clump medium, any variation in the ionization parameter arises exclusively from the density contrast. In this limit,
\begin{equation}
\label{eq:ion_ratios}
\frac{\xi_{\rm inter-clump}}{\xi_{\rm cl}}
=
\frac{\rho_{\rm cl}}{\rho_{\rm inter-clump}}
=
\frac{\alpha_m\,(1-f_V)}{(1-\alpha_m)\,f_V}\
\end{equation}
this expression should be interpreted as an effective contrast under the simplifying assumption that both phases experience comparable illumination and occupy similar characteristic radii. This yields $\rho_{\rm cl}/\rho_{\rm inter-clump}\sim 1.9\times10^{3}$ for 4U~1700$-$37 and $\sim 1.0\times10^{3}$ for XTE~J1855$-$026. Adopting a representative filling factor of $f_V\simeq0.05$, consistent with values reported for supergiant HMXB winds \citep[e.g., $f_V\approx0.04$ for Vela~X-1 and $f_V\approx0.05$ for QV~Nor]{1999ApJ...525..921S,2021MNRAS.501.5646M}, implies $\alpha_m\simeq0.99$ and $\alpha_m\simeq0.98$ for 4U~1700$-$37 and XTE~J1855$-$026, respectively.

In the literature, the clump-to-inter-clump density contrast has instead been inferred from the relative emission measures of the two \texttt{photemis} components used to model the eclipse spectra. Since the \texttt{photemis} normalization is proportional to the emission measure, $EM=\int n^2\,dV$, the ratio between the cold (clump-dominated) and hot (inter-clump-dominated) components gives
\begin{equation}
\frac{EM_{\rm cold}}{EM_{\rm hot}}
\simeq
\left(\frac{n_{\rm cl}}{n_{\rm inter-clump}}\right)^2
\frac{f_V}{1-f_V},
\end{equation}
where $f_V$ is the clump volume filling factor. For XTE~J1855$-$026, adopting $f_V\simeq0.04$--$0.05$ yields $n_{\rm cl}/n_{\rm inter-clump}\sim 180$ \citep{2022MNRAS.512..304S}. For 4U~1700$-$37, combining the measured emission-measure ratio with an independently motivated clumping factor from optical/UV studies gives $n_{\rm cl}/n_{\rm inter-clump}\sim 300$ and $f_V\simeq0.05$ \citep{10.1093/mnras/staa3956,Hainich_2020}.

To assess the impact of the adopted clumping prescription we will use both clumping prescriptions and compare the obtained results. To construct $\chi_{\rm cl}$ and $\chi_{\rm ic}$ for each of the clumping prescriptions, we used parameters collected in Table \ref{tab:system_overview}. 

\subsection{The observed $\xi$}

To apply the inference described in Sect.~\ref{sec:Single-constraint case: maximum observable ionization}, we estimate the maximum observable ionization parameter by identifying the highest-ionization emission line detected in the eclipse spectra. To infer $\xi_{\max}$ from this line, we used \texttt{XSTAR} to compute the corresponding charge-state distributions for the adopted SED (see Figs.~\ref{fig:ion_abund} and \ref{fig:mainionmatrix}). In this first-order approach, the stellar wind is assumed to be irradiated by the direct emission of the neutron star. This incident radiation field is best represented by the unabsorbed continuum derived from the observed out-of-eclipse spectra.

For 4U~1700$-$37, the out-of-eclipse spectrum can be satisfactorily described by a \texttt{powerlaw} component with a photon index of 1.35 (see \citealt{2005A&A...432..999V}). In the case of XTE~J1855$-$026, \citet{2018JApA...39....7D} reported a \texttt{powerlaw} component with a photon index of 1.12. For the sake of simplicity and to facilitate reproducibility, we adopted for both sources a spectral energy distribution (SED) represented by a \texttt{powerlaw} with a fixed photon index of 1.2. The highest-ionization emission lines detected during eclipse by \citet{2021MNRAS.501.5646M} and \citet{2022MNRAS.512..304S} are \ion{Fe}{xxv} He$\alpha$ in 4U~1700$-$37 and \ion{Fe}{xxvi} Ly$\alpha$ in XTE~J1855$-$026. Using the corresponding SEDs in \texttt{XSTAR}, these lines imply maximum ionization parameters of $\xi_{\max} \sim 2$ and $\sim 2.25$, respectively.

To apply the inference described in Sects.~\ref{sec:Two-component case: joint clump/inter-clump constraints} and \ref{sec:Separated-component case}, we directly use the $\xi_{\mathrm{cold}}$ and $\xi_{\mathrm{hot}}$ values obtained from the two \texttt{photemis} components used to model the eclipse spectra in \cite{2021MNRAS.501.5646M} and \cite{2022MNRAS.512..304S}, which we associate with the clump and inter-clump components in both systems.

\subsection{Results from the fit}

\begin{table*}
\centering
\caption{Posterior summaries for 4U~1700$-$37 and XTE~J1855$-$026.}
\label{tab:results_refac_sources_dual}

\setlength{\tabcolsep}{3.5pt}

\begin{minipage}[t]{0.49\textwidth}
\centering
\small
Clumping prescription from EM ratios

\vspace{1mm}

\begin{tabular}{lccccc}
\hline\hline
Run & $L_{\mathrm X}$ & $\dot{M}$ & $v_\infty$ & $\log_{10}\xi_{\rm pred}$ & $\log_{10}\xi_{\rm obs}$ \\
 & $(10^{35})$ & $(10^{-7})$ & $(\mathrm{km\,s^{-1}})$ & & \\
\hline
\multicolumn{6}{l}{4U~1700$-$37} \\
\hline
clumps      & $7^{+2}_{-3}$ & $190^{+90}_{-80}$ & $1900^{+300}_{-400}$ & $-1.0\pm0.1$ & $-1.0$ \\\\
inter-clump & $8.5^{+1.1}_{-1.6}$ & $26.2^{+7.7}_{-4.5}$ & $2060^{+240}_{-360}$ & $2.3\pm0.1$ & $2.38$ \\\\
max ion     & $7.1^{+2.1}_{-3.1}$ & $173^{+90}_{-77}$ & $1900^{+300}_{-400}$ & $2.0\pm0.1$ & $2.0$ \\\\
joint$_{c}$ & $7^{+2}_{-3}$ & $61^{+26}_{-23}$ & $1900^{+300}_{-400}$ & $-0.5\pm0.1$ & $-1.0$ \\
joint$_{s}$ &  &  &  & $1.9\pm0.1$ & $2.38$ \\
\hline
\multicolumn{6}{l}{XTE~J1855$-$026} \\
\hline
clumps      & $15^{+4}_{-5}$ & $3.2^{+1.6}_{-1.2}$ & $650^{+120}_{-140}$ & $0.34\pm0.1$ & $0.36$ \\\\
inter-clump & $15^{+3}_{-5}$ & $0.2\pm0.1$ & $670^{+110}_{-140}$ & $3.7\pm0.1$ & $3.7$ \\\\
max ion     & $15^{+4}_{-5}$ & $25^{+12}_{-10}$ & $650^{+120}_{-150}$ & $2.23\pm0.1$ & $2.25$ \\\\
joint$_{c}$ & $14.7^{+3.8}_{-5.5}$ & $0.8\pm0.3$ & $650^{+120}_{-140}$ & $1.0\pm0.1$ & $0.36$ \\
joint$_{s}$ &  &  &  & $3.0\pm0.1$ & $3.7$ \\
\hline
\end{tabular}
\tablefoot{ Literature density-contrast setup.}
\end{minipage}\hfill
\begin{minipage}[t]{0.49\textwidth}
\centering
\small
Clumping prescription from $\xi$ ratios

\vspace{1mm}

\begin{tabular}{lccccc}
\hline\hline
Run & $L_{\mathrm X}$ & $\dot{M}$ & $v_\infty$ & $\log_{10}\xi_{\rm pred}$ & $\log_{10}\xi_{\rm obs}$ \\
 & $(10^{35})$ & $(10^{-7})$ & $(\mathrm{km\,s^{-1}})$ & & \\
\hline
\multicolumn{6}{l}{4U~1700$-$37} \\
\hline
clumps      & $7^{+2}_{-3}$ & $150^{+80}_{-70}$ & $1900^{+300}_{-400}$ & $-1.0\pm0.1$ & $-1.0$ \\\\
inter-clump & $7^{+2}_{-3}$ & $100^{+50}_{-40}$ & $1900^{+300}_{-400}$ & $2.4\pm0.1$ & $2.38$ \\\\
max ion     & $7^{+2}_{-3}$ & $69^{+34}_{-28}$ & $1900^{+300}_{-400}$ & $2.0\pm0.1$ & $2.0$ \\\\
joint$_{c}$ & $7^{+2}_{-3}$ & $120\pm50$ & $1900^{+300}_{-400}$ & $-0.9\pm0.1$ & $-1.0$ \\
joint$_{s}$ &  &  &  & $2.3\pm0.1$ & $2.38$ \\
\hline
\multicolumn{6}{l}{XTE~J1855$-$026} \\
\hline
clumps      & $15^{+4}_{-5}$ & $3\pm1$ & $650^{+120}_{-150}$ & $0.3\pm0.1$ & $0.36$ \\\\
inter-clump & $15^{+4}_{-5}$ & $0.9^{+0.5}_{-0.4}$ & $650^{+120}_{-140}$ & $3.7\pm0.1$ & $3.7$ \\\\
max ion     & $14^{+4}_{-5}$ & $45^{+22}_{-18}$ & $650^{+120}_{-150}$ & $2.23\pm0.1$ & $2.25$ \\\\
joint$_{c}$ & $15^{+4}_{-5}$ & $1.6^{+0.7}_{-0.6}$ & $650^{+120}_{-140}$ & $0.6\pm0.1$ & $0.36$ \\
joint$_{s}$ &  &  &  & 3.4$\pm$0.1 \\
\hline
\end{tabular}

\tablefoot{Density contrast inferred from $\xi$ ratios.}

\end{minipage}
\end{table*}

Table~\ref{tab:results_refac_sources_dual} shows that the two-phase ionization structure inferred from the eclipse spectra is qualitatively recovered in both sources under both clumping prescriptions. In all separated-component runs, the low-ionization solution is associated with the clump phase, whereas the high-ionization solution is associated with the inter-clump phase. This indicates that the framework naturally reproduces the observed division between a dense, weakly ionized component and a more tenuous, highly ionized one.

A first robust result is that the inferred terminal velocities and luminosities remain comparatively stable across runs. For 4U~1700$-$37, all solutions cluster around \(v_\infty \sim 1900\)--\(2100\)~km~s\(^{-1}\), whereas for XTE~J1855$-$026 they consistently favor \(v_\infty \sim 650\)--\(700\)~km~s\(^{-1}\). The inferred \(L_{\mathrm X}\) values also remain broadly consistent within each source. By contrast, \(\dot{M}\) is more sensitive to the adopted clumping contrast. Within the present framework, this is expected because \(\dot{M}\) sets the density normalization of both the clump and inter-clump components, and is therefore the parameter most directly affected by the assumed wind clumping prescription.

A direct comparison can be made with the clumpy-wind model of \citet{2009MNRAS.398.2152D}, who also studied 4U~1700$-$37. In their work, the X-ray flare luminosity and duration distributions were modeled as the direct accretion of wind clumps within a Bondi--Hoyle framework. Their best-fitting description adopted a fraction \(f=0.75\) of the wind mass contained in clumps, together with \(\dot{M}=1.3\times10^{-6}\,M_\odot\,{\rm yr}^{-1}\) and \(v_\infty=1700\,{\rm km\,s^{-1}}\).
These values can be compared with the wind parameters obtained in our analysis for 4U~1700$-$37 from the joint fit, which is the most informative case because  it accounts for both the clump and inter-clump phases of the wind. The two clumping prescriptions explored in this analysis yield broadly similar global wind parameters in this case (see Table \ref{tab:results_refac_sources_dual}). The terminal velocity inferred in our analysis, \(v_\infty=1900\pm400\,{\rm km\,s^{-1}}\), is fully consistent with the value adopted by \citet{2009MNRAS.398.2152D}. Our mass-loss rate, \(\dot{M} \sim 6$--$7\times10^{-6}\,M_\odot\,{\rm yr}^{-1}\), is higher, but still within the same order of magnitude. The main difference concerns the inferred clump mass fraction: while \citet{2009MNRAS.398.2152D} adopted \(f=0.75\), our eclipse-ionization analysis favors larger effective values, namely \(\alpha_m\simeq0.95\) when inferred from emission-measure ratios and \(\alpha_m\simeq0.99\) when inferred from ionization-parameter ratios. When a clump mass fraction, \(\alpha_m=0.75\) is imposed, the 4U~1700$-$37 joint fit yields global wind parameters broadly
comparable to those of \citet{2009MNRAS.398.2152D}. However, the predicted clump/inter-clump ionization contrast remains smaller than
observed, and the observed \(\xi\) values cannot be reproduced. This indicates that the eclipse-ionization diagnostic favors a larger density contrast between the clump and inter-clump media than that assumed in \cite{2009MNRAS.398.2152D}.

The most informative comparison is provided by the joint fits, since these require a single set of global wind--radiation parameters to reproduce both ionization components simultaneously. In this sense, the joint scenario constitutes a more stringent test than the separated runs: while the latter verify that each phase can be matched independently, the former probes whether both phases can be understood self-consistently within the same physical configuration.

From this perspective, the prescription based on \(\xi\) ratios is favored. For XTE~J1855$-$026, both prescriptions recover the qualitative low-/high-ionization split, but the \(\xi\)-ratio prescription yields a more satisfactory joint solution, with predicted values closer to the observed pair of ionization states. For 4U~1700$-$37, the same tendency is present and is even clearer: the \(\xi\)-ratio prescription produces a joint solution in which both the low- and high-ionization components are reproduced more closely than in the EM-ratio case. Thus, although both prescriptions preserve the same qualitative picture, the \(\xi\)-ratio setup provides the more self-consistent global explanation of the eclipse spectra. This result suggests that a more extreme density contrast between the clump and inter-clump contributions favors the observed scenario. The large effective density contrasts inferred here are consistent with the broader picture of clumped irradiated winds, in which dense structures enhance recombination and reduce the impact of X-ray overionization \citep{2018A&A...620A.150K}.

The full set of convergence diagnostics and prior-boundary accumulation fractions is given in Appendix~\ref{app:mcmc_diag} (Tables~\ref{tab:diag_boundary_lit} and \ref{tab:diag_boundary_xi}). All runs show \(\hat{R}\simeq 1.00\) and effective sample sizes of order \(10^{4}\) or higher, indicating stable posterior sampling. At the same time, the strong accumulation of posterior mass at the lower prior bound of \(\dot{M}\) shows that the inference remains partly prior-limited. Therefore, the most robust outcome of this analysis is not the precise value of each individual parameter, but rather the identification of a preferred wind--radiation regime in which the observed ionization contrast between the two plasma phases is more naturally reproduced by large clump-to-inter-clump density ratios.

\subsection{Discussion on the smooth case}
For completeness, we also performed the Bayesian inference for the case of a smooth wind, using the maximum observed ionization as the only constraint. The resulting smooth-wind solutions are summarized in Table~\ref{tab:smooth_maxion}.

\begin{table}
\caption{Smooth-wind solutions constrained by the maximum observed ionization.}
\label{tab:smooth_maxion}
\centering

\begin{tabular}{lcc}
\hline\hline
Source & 4U~1700$-$37 & XTE~J1855$-$026 \\
\hline
$L_{\mathrm X}\,(10^{35}\,\mathrm{erg\,s^{-1}})$              & $9.4^{+0.5}_{-0.9}$     & $15^{+4}_{-5}$ \\\\
$\dot{M}\,(10^{-7}\,M_\odot\,\mathrm{yr}^{-1})$               & $22^{+3}_{-1}$          & $5.6^{+2.8}_{-2.2}$ \\\\
$v_\infty\,(\mathrm{km\,s^{-1}})$                             & $2240^{+120}_{-210}$    & $650^{+120}_{-150}$ \\\\
$\log_{10}\xi_{\rm pred}$                                     & $1.80\pm0.06$           & $2.23\pm0.1$ \\\\
$\log_{10}\xi_{\rm obs}$                                      & $2.0$                   & $2.25$ \\
\hline
\end{tabular}
\end{table}

Although in both sources the predicted ionization parameter remains slightly below the adopted observational value, with $\log_{10}\xi_{\rm pred} \simeq 1.8$ compared to $\log_{10}\xi_{\rm obs} = 2.0$ in 4U~1700$-$37, and $\log_{10}\xi_{\rm pred} \simeq 2.23$ versus $\log_{10}\xi_{\rm obs} = 2.25$ in XTE~J1855$-$026, in the latter source the smooth-wind case may still be regarded as broadly consistent with the observed ionization level. The discrepancy is more pronounced in 4U~1700$-$37, where the maximum ionization reached in the smooth-wind approximation remains farther from the observed value and is reproduced more naturally within the clumpy scenario.

\section{Conclusions}
\label{sec:conclusions}

We have developed a phase-dependent framework that combines a CAK-like wind prescription, a two-component medium (clumps plus inter-clump gas), and \texttt{XSTAR}-based ionization calculations to construct 3D ionization maps in HMXBs and to quantify the orbital-phase dependence of their observable ionic structure. Our main conclusions can be summarized as follows:

\begin{itemize}
    \item Even within this first-order approximation, the method recovers a substantial amount of physically relevant information. By combining wind structure, orbital geometry, and photoionization balance in a computationally efficient framework, it provides a direct and interpretable connection between phase-resolved spectroscopic diagnostics and the three-dimensional ionization structure of irradiated HMXB winds.\\

    \item The observable ionic structure depends strongly on both clumping and orbital geometry. The clump mass fraction, the spatial distribution of clumps, inclination, eccentricity, eclipse occultation, and donor-star X-ray shadowing all reshape the regions where different ions are preferentially found and modulate their visibility with orbital phase. Eclipse phases are particularly informative, because the direct compact-object continuum is suppressed and the circumstellar ionization structure can be probed more cleanly.\\

    \item The applications to 4U~1700$-$37 and XTE~J1855$-$026 support a picture in which low-ionization plasma is predominantly associated with the clump phase, whereas highly ionized plasma is preferentially linked to the inter-clump medium. 
    Results suggest that the data favor large effective density contrasts between clumps and the inter-clump medium.\\
\end{itemize}

This is a first-order study: we focus on diagnostics of the spatial ionization structure and do not yet include full radiative-transfer effects in the predicted emergent line fluxes. A natural next step is to couple the present framework to line-transfer calculations and apply it to phase-resolved eclipse spectroscopy. This approach will be particularly relevant in the era of new X-ray observatories, such as \textit{Athena} \citep{Cruise2025} and \textit{XRISM} \citep{2022IJMPD..3130001T}, where improved spectral resolution and sensitivity will enable tighter constraints on clumping and orbital structure.

\begin{acknowledgements}
JJRR, JMT, JPV and GSF acknowledge the financial support from the MICIU/AEI/10.13039/501100011033 with funding from the European Union (FEDER). Project (NewAthena24-UA), reference PID2024-155779OB-C33. RB acknowledges supported by NASA under award number 80GSFC24M0006.
\end{acknowledgements}

\bibliographystyle{aa}
\bibliography{bib.bib} 

\begin{appendix}
\section{Detailed wind prescription calculations}
\label{app:calc_xi}

In this Appendix we show how the factorized form of the ionization parameter introduced in Sect.~4 follows from the wind parametrization adopted in Sect.~2.

We start from the standard definition of the ionization parameter,
\begin{equation}
\xi = \frac{L_{\rm X}}{n\,r^2},
\end{equation}
where \(L_{\rm X}\) is the X-ray luminosity, \(n\) is the local particle number density, and \(r\) is the distance from the plasma element to the X-ray source. Using \(n=\rho/(\mu m_{\rm p})\), this can be written as
\begin{equation}
\xi = \frac{L_{\rm X}\,\mu m_{\rm p}}{\rho\,r^2}.
\label{eq:app_xi_rho}
\end{equation}

The smooth wind density is described by the CAK-like prescription adopted in Sect.~2,
\begin{equation}
\rho(d)=\frac{\dot M}{4\pi d^2 v(d)},
\end{equation}
with velocity law
\begin{equation}
v(d)=v_\infty\left(1-\frac{R_\star}{d}\right)^\beta.
\end{equation}
Here \(d\) denotes the distance from the donor-star centre. In Sect.~4 we express distances in units of \(R_\star\), so that \(d\rightarrow d/R_\star\) and \(r\rightarrow r/R_\star\). With this notation,
\begin{equation}
v(d)=v_\infty\left(1-\frac{1}{d}\right)^\beta,
\end{equation}
and the smooth density becomes
\begin{equation}
\rho(d)\propto \frac{\dot M}{d^2\,v_\infty\left(1-\frac{1}{d}\right)^\beta}.
\label{eq:app_rho_smooth}
\end{equation}

The two-component description introduced in Sect.~2 separates the wind into clump and inter-clump phases. From the definition of the clump mass fraction,
\begin{equation}
\alpha_m = \frac{f_V \rho_{\rm cl}}{\rho},
\end{equation}
the clump density is
\begin{equation}
\rho_{\rm cl}=\frac{\alpha_m}{f_V}\,\rho.
\label{eq:app_rho_cl}
\end{equation}
Likewise, using mass conservation,
\begin{equation}
\rho = f_V\rho_{\rm cl} + (1-f_V)\rho_{\rm ic},
\end{equation}
the inter-clump density can be written as
\begin{equation}
\rho_{\rm ic}=\frac{1-\alpha_m}{1-f_V}\,\rho.
\label{eq:app_rho_ic}
\end{equation}

Substituting Eq.~(\ref{eq:app_rho_cl}) into Eq.~(\ref{eq:app_xi_rho}) gives the ionization parameter in the clump phase,
\begin{equation}
\xi_{\rm cl}(d,r)
=
\frac{L_{\rm X}\mu m_{\rm p}}{\rho_{\rm cl}\,r^2}
=
\frac{f_V(d)}{\alpha_m(d)}
\frac{L_{\rm X}\mu m_{\rm p}}{\rho(d)\,r^2}.
\end{equation}
Replacing \(\rho(d)\) with Eq.~(\ref{eq:app_rho_smooth}) yields
\begin{equation}
\xi_{\rm cl}(d,r)
=
\xi_0\,
\frac{f_V(d)}{\alpha_m(d)}
\left(1-\frac{1}{d}\right)^\beta
\left(\frac{d}{r}\right)^2,
\end{equation}
where all global wind--radiation dependence has been grouped into
\begin{equation}
\xi_0=\frac{L_{\rm X}\mu m_{\rm p}v_\infty}{\dot M},
\end{equation}
up to the constant geometrical normalization adopted in the main text.

Therefore,
\begin{equation}
\xi_{\rm cl}(d,r)=\xi_0\,\chi_{\rm cl}(d,r),
\end{equation}
with
\begin{equation}
\chi_{\rm cl}(d,r)=
\frac{f_V(d)}{\alpha_m(d)}
\left(1-\frac{1}{d}\right)^\beta
\left(\frac{d}{r}\right)^2.
\end{equation}

Proceeding in the same way for the inter-clump phase, Eq.~(\ref{eq:app_rho_ic}) gives
\begin{equation}
\xi_{\rm ic}(d,r)
=
\frac{L_{\rm X}\mu m_{\rm p}}{\rho_{\rm ic}\,r^2}
=
\frac{1-f_V(d)}{1-\alpha_m(d)}
\frac{L_{\rm X}\mu m_{\rm p}}{\rho(d)\,r^2},
\end{equation}
and therefore
\begin{equation}
\xi_{\rm ic}(d,r)=\xi_0\,\chi_{\rm ic}(d,r),
\end{equation}
with
\begin{equation}
\chi_{\rm ic}(d,r)=
\frac{1-f_V(d)}{1-\alpha_m(d)}
\left(1-\frac{1}{d}\right)^\beta
\left(\frac{d}{r}\right)^2.
\end{equation}

This directly leads to the factorized form used in Sect.~4,
\begin{equation}
\xi(d,r)=\xi_0\,\chi(d,r),
\end{equation}
where \(\chi\) is a dimensionless geometric--kinematic factor that depends on the location in the wind and on whether the clump or inter-clump phase is considered. In this formulation, the dominant global scaling is shared by both phases and is controlled by
\begin{equation}
\xi_0 \propto \frac{L_{\rm X}v_\infty}{\dot M},
\end{equation}
while all local differences are encoded in \(\chi_{\rm cl}\) and \(\chi_{\rm ic}\).

\section{MCMC Diagnostics}
\label{app:mcmc_diag}

\begin{table*}
\centering
\caption{Combined MCMC diagnostics and boundary-accumulation fractions (run order: clumps, inter-clump, max ion, joint) for the literature density-contrast setup.}
\label{tab:diag_boundary_lit}
\small
\setlength{\tabcolsep}{4.5pt}
\renewcommand{\arraystretch}{1.10}
\begin{tabular}{lccccc}
\hline\hline
Run & ESS$_{\rm bulk}$ min--max & ESS$_{\rm tail}$ min--max &
$f_{\rm low}(\dot{M})$ & $f_{\rm up}(L_{\mathrm X})$ & $f_{\rm low}(v_\infty)$ \\
\hline
\multicolumn{6}{l}{4U~1700$-$37}\\
\hline
clumps  & 13010--41557 &  9932--29283 & 1.00 & 0.02 & 0.00 \\
inter-clump  & 20130--34400 & 14403--24829 & 1.00 & 0.08 & 0.00 \\
max ion & 13807--38293 &  9872--26897 & 1.00 & 0.02 & 0.00 \\
joint   & 18607--33588 & 12596--25910 & 1.00 & 0.24 & 0.00 \\
\hline
\multicolumn{6}{l}{XTE~J1855$-$026}\\
\hline
clumps  & 15695--41845 & 16670--32153 & 1.00 & 0.80 & 0.20 \\
inter-clump  & 17456--41332 & 12873--27761 & 1.00 & 0.88 & 0.18 \\
max ion & 14502--38000 & 14226--30233 & 1.00 & 0.81 & 0.21 \\
joint   & 11959--38452 & 11794--31005 & 1.00 & 0.80 & 0.20 \\
\hline
\end{tabular}
\tablefoot{ESS ranges are computed across all sampled parameters in each run (including $\log_{10}\xi_{\rm pred}$ where applicable). For all sampled parameters in all runs, $\hat{R}=1.00$ (within reported precision). Boundary fractions are rounded to two decimals.}
\end{table*}

\begin{table*}
\centering
\caption{Combined MCMC diagnostics and boundary-accumulation fractions (run order: clumps, inter-clump, max ion, joint) inferred from $\xi$ ratios.}
\label{tab:diag_boundary_xi}
\small
\setlength{\tabcolsep}{4.5pt}
\renewcommand{\arraystretch}{1.10}
\begin{tabular}{lccccc}
\hline\hline
Run & ESS$_{\rm bulk}$ min--max & ESS$_{\rm tail}$ min--max &
$f_{\rm low}(\dot{M})$ & $f_{\rm up}(L_{\mathrm X})$ & $f_{\rm low}(v_\infty)$ \\
\hline
\multicolumn{6}{l}{4U~1700$-$37}\\
\hline
clumps  & 13693--41892 & 11429--29485 & 1.00 & 0.02 & 0.00 \\
inter-clump  & 13302--40419 & 10352--28601 & 1.00 & 0.02 & 0.00 \\
max ion & 20696--32479 & 14399--25887 & 1.00 & 0.64 & 0.00 \\
joint   & 11659--42159 &  9645--30871 & 1.00 & 0.02 & 0.00 \\
\hline
\multicolumn{6}{l}{XTE~J1855$-$026}\\
\hline
clumps  & 14272--40193 & 15583--32033 & 1.00 & 0.80 & 0.20 \\
inter-clump  & 13735--38526 & 13904--30704 & 1.00 & 0.81 & 0.20 \\
max ion & 20973--33949 & 14307--25442 & 1.00 & 1.00 & 0.00 \\
joint   & 11904--40975 & 12614--31755 & 1.00 & 0.81 & 0.20 \\
\hline
\end{tabular}
\tablefoot{ESS ranges are computed across all sampled parameters in each run (including $\log_{10}\xi_{\rm pred}$ and, for the joint runs, both phase-specific $\mu$ parameters). For all sampled parameters in all runs, $\hat{R}=1.00$ (within reported precision). Boundary fractions are rounded to two decimals.}
\end{table*}

\end{appendix}

\onecolumn

\end{document}